# Mixed platinum and zirconia powder as electrocatalyst for hydrogen evolution and oxidation reaction

Simone Minelli[1], Manuel Iozzia[2], Claudio Piazzoni[2], Alberto Vertova[*,1,3], Cristina Lenardi[2], Alessandro Minguzzi[1,3,4]

1 Dipartimento di Chimica, Università degli Studi di Milano, Via Golgi 19, 20133 - Milano, Italy.
2 Dipartimento di Fisica, Università degli Studi di Milano, Via Golgi 19, 20133 ` Milano, Italy.
3 INSTM Consorzio Interuniversitario Nazionale per la Scienza e Tecnologia dei Materiali, Via G. Giusti 9, 50121 Firenze, Italy
4 Dipartimento di Energia, Politecnico di Milano, Via Lambruschini 4a, 20156 Milano, Italy

*Corresponding author: alberto.vertova@unimi.it

## Abstract

The development of efficient and cost-effective bifunctional electrocatalysts for hydrogen-related technologies is crucial for the transition to sustainable energy. In this work, we present Pt@$ZrO_2$ as a high-performance ceramic-based electrocatalyst for the hydrogen evolution reaction (HER) and hydrogen oxidation reaction (HOR), demonstrating superior activity and stability in alkaline conditions. Compared to Pd@$ZrO_2$, previously investigated in our earlier work, Pt@$ZrO_2$ exhibits significantly lower charge transfer resistance and enhanced HER/HOR kinetics, as confirmed by electrochemical impedance spectroscopy. The synergistic interaction between platinum and $ZrO_2$ stabilizes Pt in a partially oxidized state, facilitating the Volmer step and improving charge transfer kinetics. Cyclic voltammetry further confirms the remarkable electrochemical stability of Pt@$ZrO_2$ under stress conditions. Importantly, Pt@$ZrO_2$ offers a cost-effective alternative to pure platinum by reducing noble metal loading while maintaining excellent catalytic performance. These findings establish Pt@$ZrO_2$ as a better alternative to Pd@$ZrO_2$ and a promising candidate for next-generation hydrogen energy applications.

# Introduction

The escalating global energy demand, primarily driven by rapid industrialization and population growth, has intensified the need for sustainable and environmentally friendly alternatives to fossil fuels [1,2]. Among the various emerging solutions, hydrogen has gained significant attention as a clean energy carrier thanks to its high energy density and the absence of pollutant emissions during its use [3].

Water electrolysis stands out as a promising method for generating green hydrogen using renewable electricity. We have recently demonstrated not only the usage of copper oxides as promising photocathode for water splitting reaction [4,5], but also the possibility of using the hydrogen oxidation and reduction reaction as effective method to purify a gaseous flux containing $H_2$ [6].

In particular, Anion Exchange Membrane Water Electrolysis (AEMWE) has emerged as a compelling technology due to its unique advantages [7,8]. Operating under alkaline conditions, AEMWE systems allow the use of non-precious metal catalysts and cost-effective components, while maintaining high ionic conductivity and system efficiency [9–11]. However, the use of Platinum Group Metals (PGMs) as catalysts for the Hydrogen Evolution Reaction (HER) remains a significant challenge for AEMWE due to their high cost and limited availability [12,13]. As highlighted in the life cycle analysis (LCA) study by our group [14], the amount of platinum used in AEMWE systems has a substantial impact on both the capital expenditures (CAPEX) and the environmental footprint of the hydrogen production process. Consequently, reducing the amount of PGMs while maintaining or improving catalytic efficiency is a critical focus of ongoing research [15].

To address these challenges, composite electrocatalysts have emerged as a promising solution in general for the hydrogen and oxygen electrode and photoelectrode reactions and in particular for water electrolysis [16–18]. By combining the catalytic properties of PGMs with non-precious, oxyphilic materials, such as ceria ($CeO_2$), researchers have shown that it is possible to enhance HER activity and reduce overpotential, thereby improving the efficiency of hydrogen production in alkaline media [19–21]. These improvements are mainly attributed to both the dimension of the particles forming the electrodes, as demonstrated for sol-gel based nanopowders used in electrochemical application [22], and to the synergistic interactions between the two components, which facilitate water adsorption, lower the energy barriers for hydrogen evolution, and enhance catalyst stability, as demonstrated in several studies [23–26]. These interactions are particularly beneficial in alkaline environments, where the HER mechanism differs significantly from that in acidic media due to the lower availability of protons ($H^+$), following the Volmer–Heyrovsky mechanism, where water molecules are adsorbed on the catalyst surface and split into hydrogen atoms and hydroxide ions [27].

While CeO$_2$ has proven effective in boosting the performance of Pd-based catalysts for HER, its relatively high cost and limited availability compared to other materials have prompted the search for alternative, more cost-effective oxyphilic supports. In this regard, zirconia (ZrO$_2$) presents a promising alternative. ZrO$_2$ is a non-toxic, eco-friendly ceramic material known for its mechanical strength, biocompatibility, and stability under harsh conditions, making it an ideal candidate for electrocatalytic applications [28–30]. Recent studies have shown that ZrO$_2$ can similarly enhance the catalytic properties of PGMs in alkaline HER, providing a more sustainable solution for large-scale hydrogen production [31,32].

The goal of this research is to develop a highly active Pt@ZrO$_2$ composite electrocatalyst for the HER in alkaline environments. Building on the work of our group [31], who demonstrated the effectiveness of Pd@ZrO$_2$ composites, this study aims to replace Pd with Pt, as Pd is less active than Pt for HER in alkaline media [33–36]. A key advantage of this approach is the use of Ion Beam Sputtering Deposition (IBSD) to synthesize the composite material directly onto the substrate. IBSD allows for precise control of catalyst composition and structure, ensuring high reproducibility and adhesion of the composite layer. Moreover, by depositing the catalyst directly on Anion Exchange Membranes (AEMs), the need for separate catalyst layers is eliminated, reducing material costs and simplifying the manufacturing process. This approach not only contributes to reducing the platinum load but also holds promise for enhancing the scalability and cost-effectiveness of AEMWE systems [14].

This work presents the synthesis, characterization, and electrochemical performance of Pt@ZrO$_2$ composite electrocatalysts, demonstrating their potential for efficient hydrogen evolution in alkaline electrolysers.

# Experimental section

*Composite Sputtering Target Preparation*

Pt@ZrO$_2$ composite electrocatalyst materials were synthesized using the Ion Beam Sputtering Deposition (IBSD) technique, a physical vapor deposition (PVD) technique that allows precise control over the properties of the deposited film. By tuning parameters such as ion species, energy, and incidence angle, it is possible to influence key characteristics like density, adhesion, surface roughness, and electrical resistivity [37].

In this study, IBSD was employed with a specifically engineered Pt/Zr composite target to achieve a uniform co-deposition of both elements in controlled proportions as schematically depicted in Figure 1. The IBSD system features a Kaufman ion beam source operating within a 100 dm³ vacuum chamber, maintaining a base pressure in the $10^{-4}$ Pa range. Argon ions (Ar$^+$) serve as the primary sputtering species. The composite target, positioned at a 30° incidence angle to the ion beam, consists of a 100 x 133 x 0.3 mm Pt foil overlaid with a Zr mesh composed of 0.2 mm diameter Zr wire. The mesh is secured onto the Pt foil using an aluminum frame with grooves (0.6 mm width, 1 mm pitch), creating an exposed area with approximately 36% Pt and 64% Zr. This configuration was designed to achieve a final Pt/Zr ratio of 50:50, considering the sputtering yield of each element [38,39]. Deposition thickness was continuously monitored in situ via a quartz microbalance placed within the sputtered area. The Pt@$ZrO_2$ films were deposited onto 20 x 10 mm fluorine-doped tin oxide (FTO) substrates (Sigma-Aldrich, sheet resistance ~7 Ω/sq). A masking technique was applied to restrict the coated region to approximately 10 x 10 mm of the conductive surface. Film thickness measurements were performed on FTO substrates using stylus profilometry. Four electrodes with different platinum thicknesses (25, 50, 100, and 200 nm) were fabricated. Electrochemical characterizations were performed on the Pt@$ZrO_2$ 50 nm sample, as this thickness was previously identified as the most promising candidate in our earlier study on Pd@$ZrO_2$-based electrodes [31].

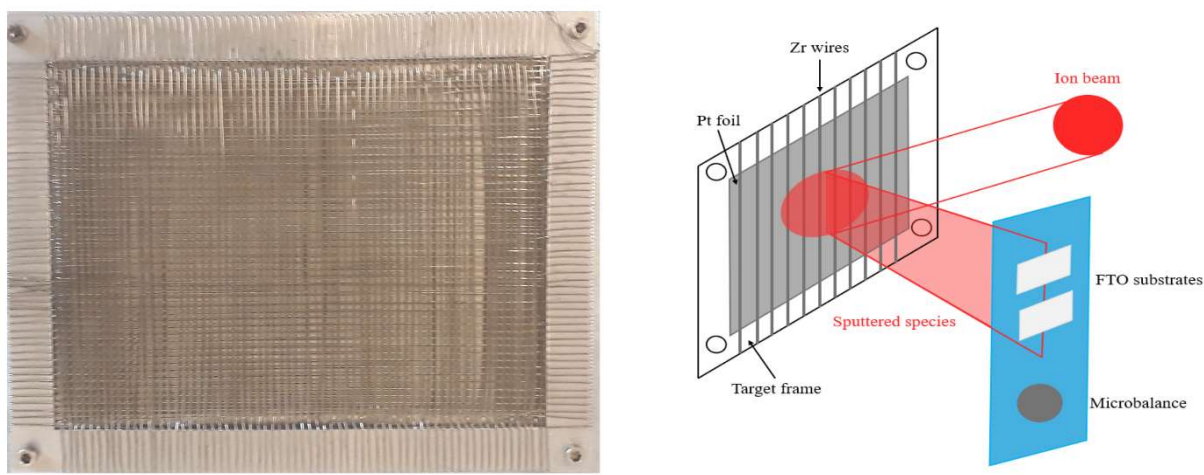

*Figure 1: Pt and Zr composite sputtering target composed of a Platinum foil, an aluminium frame and a Zirconium wire with an exposed area of 36% Pt and 64% Zr (left) and schematic drawing of the deposition geometry (right).*

## Profilometry

A KLA-Tencor stylus profilometer, featuring a low-force capacitive sensor and a 2 μm radius tip, was employed to determine the thickness of the Pt@$ZrO_2$ coating on FTO substrates.

## X-ray Photoelectron Spectroscopy (XPS) Characterization

The electrocatalytic composite films obtained via IBSD were analyzed by X-ray Photoelectron Spectroscopy (XPS) to assess: i) the elemental composition of the IBSD-deposited films and ii) the oxidation states of Pt and Zr. XPS measurements were conducted at room temperature using a custom-built ultra-high vacuum (UHV) system, featuring a 150 mm hemispherical analyzer PHOIBOS150 from SPECS Gmb and a non-monochromatized Al K$\alpha$ X-ray source. High-resolution spectra were collected with a pass energy of 20 eV, achieving an overall energy resolution of approximately 1 eV (Full Width at Half Maximum). During the measurements, the chamber pressure was maintained in the low $10^{-10}$ mbar range. Peaks were fitted after Shirley background subtraction and at.% of elemental compositions were extracted from peak area ratios after correction by Scofield relative sensitivity factors (Pt=15.45 and Zr=7.04) [40].

*Electrochemical Characterization*

The electrochemical performance of the electrodes prepared by IBSD was evaluated using cyclic voltammetry (CV) and Electrochemical Impedance Spectroscopy (EIS) in a 3-electrode setup. The electrochemical tests were carried out in two different KOH solutions: 0.1 M and 1 M (Sigma-Aldrich reagent grade pellets) thoroughly purged with $N_2$ to eliminate $O_2$ and $CO_2$. A Pt foil was used as the counter electrode (CE), and a Saturated Calomel Electrode (SCE) was employed as the reference electrode, with a double bridge filled with 0.5 M $KNO_3$ solution.

For CV measurements, the potential range was selected from -0.05 $V_{RHE}$ to 1.4 $V_{RHE}$ to avoid the formation of Pt oxides and the Oxygen Evolution Reaction (OER). The scan rate was 20 mV/s, except during the "stress test" cycles, where a scan rate of 1 V/s was used for 5000 cycles. For EIS measurements, an amplitude of 10 mV and a frequency range from 100 kHz to 0.1 Hz was applied at a potential of -0.275 $V_{RHE}$. Prior to EIS, the electrode was polarized from 0 to -0.275 $V_{RHE}$ with a scan rate of 10 mV/s.

The electrochemical characterization was carried out using a Solartron 1287 potentiostat / galvanostat coupled with a Solartron 1260 Frequency Response Analyzer (FRA). Data collection and analysis were performed with CorrWare® and Zplot® software, while EIS data were processed using ZView® software.

# Results and discussion

*XPS*

The XPS analysis provides detailed insights into the elemental composition and the chemical state of the films within the first few Ångströms near the surface. To verify if the elemental composition aligns with the expected Pt/Zr ratio of 1:1, XPS measurements were conducted on all Pt@$ZrO_2$ samples with thicknesses of 25, 50, 100, and 200 nm. As reported in Table 1, the results show that Pt and Zr concentrations are generally consistent with the values predicted from the theoretical sputtering yield. The only exception is the 25 nm-thick sample. This deviation is likely due to the presence of zirconium oxide on the surface of the Zr wire within the composite target, which may significantly influence the initial stages of IBSD as $ZrO_2$ has a lower sputtering yield compared to Zr [39]. The outer oxide layer is formed with target exposition to air and is progressively removed during deposition; the result is that for higher thickness (longer deposition time) its overall effect is reduced. For future applications of the IBSD technique, it will be crucial to optimize the deposition parameters and to consider alternative strategies for composite target preparation.

*Table 1: Atomic concentrations of Pt and Zr with varying Pt@ZrO2 thicknesses (25, 50, 100, and 200 nm)*

| Sample thickness | Zr (at%) | Pt (at%) |
| --- | --- | --- |
| 25 nm | 38 | 62 |
| 50 nm | 45 | 55 |
| 100 nm | 41 | 59 |
| 200 nm | 50 | 50 |
| Expected from sputtering yield | 50 | 50 |

To assess the oxidation states of Pt and Zr, which are indicators of the synergistic interaction between Pt and $ZrO_2$, detailed scans were acquired for the Zr 3d and Pt 4f energy regions (see Figure S1 in Supporting Information). The analysis revealed that Zr is fully oxidized in the form of Zr(IV) ($ZrO_2$), with the characteristic peak at 182.1 eV, and no Zr(0) peak is present at 178.9 eV [41,42] as shown in Figure 2A. As regard platinum, partial oxidation to PtO and $PtO_2$ is observed (see in Figure 2B). The majority of Pt is in the metallic state (Pt(0)), identified by the peak centred at 71 eV for the $4f_{7/2}$ level. The remaining platinum is in the oxidized forms: as PtO, identified by the peak at 72.4 eV, and as $PtO_2$, observed at 74 eV [43,44]. The partial oxidation of the noble metal is the result of the synergistic interaction with the oxyphilic element ($ZrO_2$) that maintains the noble metal partially oxidized by sharing with Pt its abundant oxygen atoms. The quantitative evaluation of the oxidized platinum was performed by fitting the peaks using Voigt line-shapes. The quality of the fit was evaluated using the Abbe criterion [45]. The results, reported in Table 2, show that in all investigated samples, the majority of Pt remains in its metallic form, with values ranging from 67% to 72%. The oxidized species, PtO and $PtO_2$, are consistently present in lower amounts, with PtO varying between 22% and 26%, and $PtO_2$ between 6% and 9%. These values confirm that platinum is only partially oxidized, as expected due to the interaction with the zirconia matrix. Notably, the relative amounts of oxidized species remain fairly constant across the different samples, suggesting that the platinum oxidation state is primarily dictated by the local chemical environment rather than the nominal film thickness, which supports the hypothesis that ZrO₂ plays an active role in stabilizing oxidized Pt species on the surface.

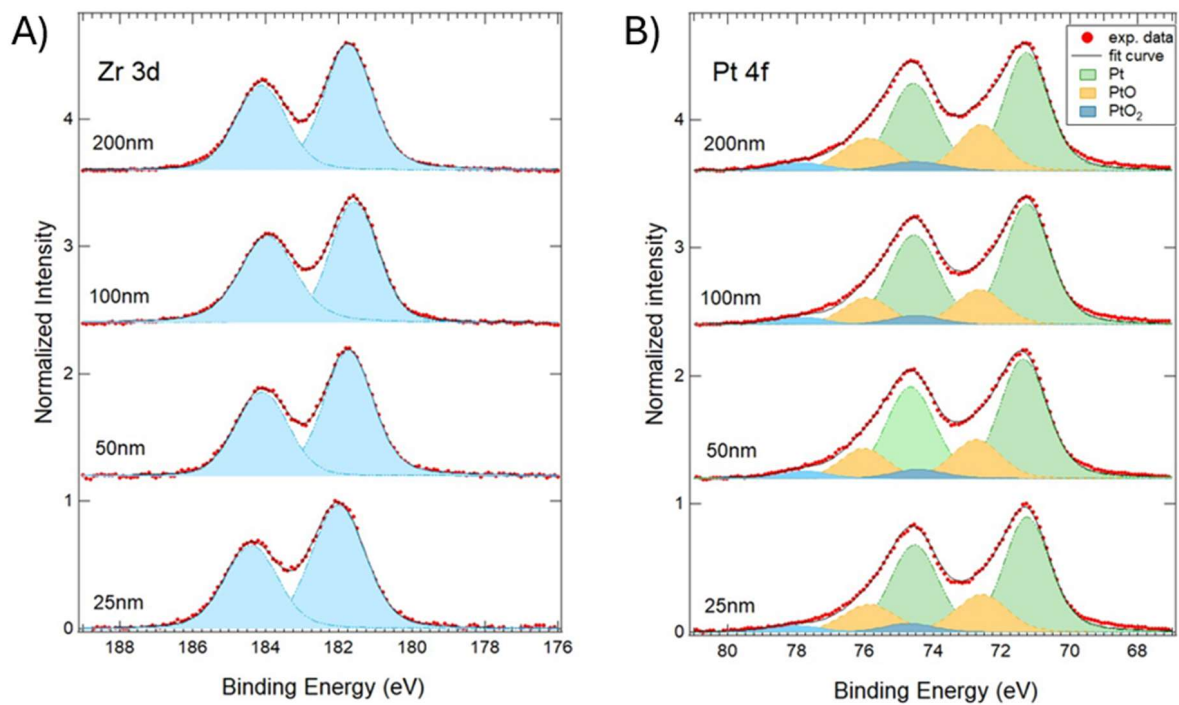

*Figure 2: Normalized XPS spectra of pristine samples at (A) Zr 3d and (B) Pt 4f energy regions for the four different thicknesses. Metallic and oxide species resulting from fitting procedure of the normalized XPS spectra are also reported.*

*Table 2: Relative abundance of metallic Pt, PtO, and PtO₂ species in Pt@ZrO₂ electrodes with different Pt thicknesses (25, 50, 100, and 200 nm), obtained from quantitative XPS peak fitting.*

| Sample thickness | Pt(0) (%) | PtO (%) | PtO$_2$ (%) |
|---|---|---|---|
| 25 nm | 70 | 24 | 6 |
| 50 nm | 72 | 23 | 6 |
| 100 nm | 69 | 22 | 9 |
| 200 nm | 67 | 26 | 7 |

Unlike in our previous study [31], where the Pd@ZrO₂ sample with a thickness of 50 nm exhibited a higher degree of palladium oxidation compared to the other thicknesses, no significant trend in platinum oxidation is observed among the Pt@ZrO₂ samples analysed in this work. For this reason, the Pt@ZrO₂ sample with 50 nm thickness was selected for direct comparison with the previously studied Pd@ZrO₂ system, enabling a more consistent evaluation of the influence of the noble metal on the interfacial chemistry with zirconia.

*Cyclic voltammetry (CV)*

Cyclic voltammetry measurements were performed in 0.1M and 1M KOH solutions to assess the electrochemical behaviour of the bifunctional electrocatalysts.

To verify whether the presence of $ZrO_2$ in direct interaction with Pt effectively decreases the overpotential of HER in an alkaline environment, the Pt@$ZrO_2$/FTO 50 nm sample was compared with a 50 nm thick Pt/FTO (without $ZrO_2$) sample deposited via IBSD. The obtained CVs results are compared in terms of mass activity, current divided for Pt calculated amount. From the comparison of the Pt@$ZrO_2$/FTO 50 nm and Pt/FTO 50 nm CV characterizations in both the selected KOH solution, reported in Figure 3, the synergistic interaction between Pt and $ZrO_2$ is evidenced. The presence of $ZrO_2$, intimately mixed with Pt, increases all the electrochemical properties of the noble metal. In the HER, the mass activity of Pt in synergistic interaction with $ZrO_2$ increases more than 3 times in 0.1M KOH (pH=13) and 6 times in case of 1M KOH (pH=14). In addition, as highlighted by Gao et al. [46] for $CeO_2$, the presence of an oxyphilic material enhances both the HER and HOR currents as KOH concentration increases, due to the synergistic effect between $CeO_2$ and Pd. A similar effect has been observed in our previous work with Pd@$ZrO_2$ [31], and is also evident in the Pt@$ZrO_2$ composite electrodes of this current work, where the interaction between the noble metal and the ceramic compound contributes to an improvement in both HER and HOR performance [47]. Specifically, the HER shows a pronounced sharp peak in the current (see Figure 3), accompanied by a significant HOR signal during the reverse cycle, followed by a noticeable plateau associated with the growth of the metal oxide. The subsequent reduction of the oxide in the reverse cycle is reflected in a distinct "bell-shaped" peak.

Having compared the performance of Pt$ZrO_2$ with pure Pt, the next step is to examine the behaviour of Pt$ZrO_2$ in comparison to Pd$ZrO_2$, as shown in Figure 4. The results demonstrated that Pt@$ZrO_2$ exhibited superior electrochemical properties compared to Pd@$ZrO_2$ in both alkaline solutions. In particular, Pt@$ZrO_2$ exhibited a higher current density at -0.05 V vs. RHE, along with more pronounced electrochemical features. These included enhanced reduction of $PtO_x$ (0.8–0.5 V vs. RHE), H-UPD (0.4–0.1 V vs. RHE), $H_2$ desorption (0.1–0.5 V vs. RHE), and Pt oxidation (0.8–1.4 V vs. RHE), suggesting a greater number of active sites available for catalytic reactions. The obtained result is in line with literature evidence that report a higher exchange current density of Pt respect of Pd [48,49].

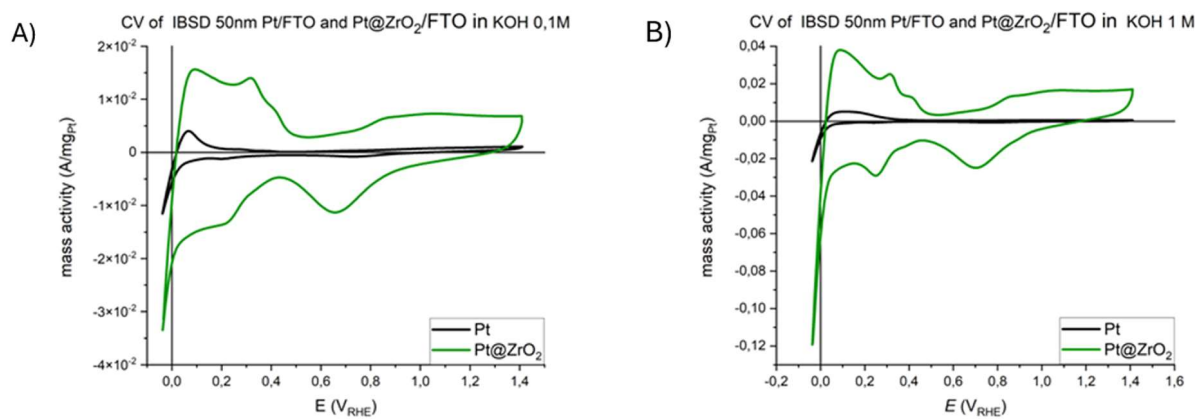

*Figure 3: Comparison between CVs recorded with 50 nm Pt/FTO (black curve) and 50 nm Pt@ZrO$_2$/FTO (green curve) electrodes in 0.1 M KOH (A) and in 1 M KOH (B)*

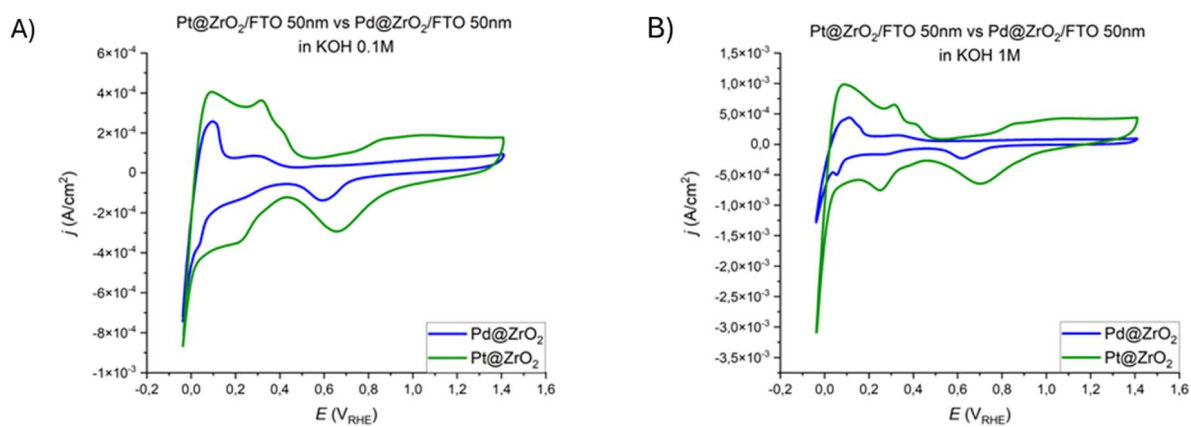

*Figure 4: Comparison of the CVs recorded with 50 nm Pd@ZrO$_2$/FTO and 50nm Pt@ZrO$_2$/FTO electrodes in (A) 0.1 M KOH 0.1M and (B) 1 M KOH solutions*

## Stability test

To assess the stability of the Pt@ZrO$_2$/FTO composite electrocatalyst, a stress test consisting of 5000 CV cycles was performed in both 0.1 M and 1 M KOH solutions. The CV curves recorded before and after the stress test, shown in Figure 5, reveal different behaviours depending on the electrolyte concentration. In 0.1 M KOH, the electrochemical properties remained unchanged throughout the stress test, indicating that the composite electrocatalyst retained its electrochemical stability. However, in 1 M KOH, notable changes were observed. While the HER current density at -0.05 V vs. RHE showed only a slight decrease from 3 mA/cm² to 2.7 mA/cm², and the HOR peak at 0.05 V vs. RHE remained unaffected, other electrochemical features underwent significant modifications. A possible explanation is the partial loss of ZrO$_2$ in more alkaline solution during the 5000 CV cycles, the current decreasing points to an increase in Pt/ZrO$_2$ ration, which, despite altering some electrochemical characteristics, did not impact the HER and HOR activity of the composite electrocatalyst.

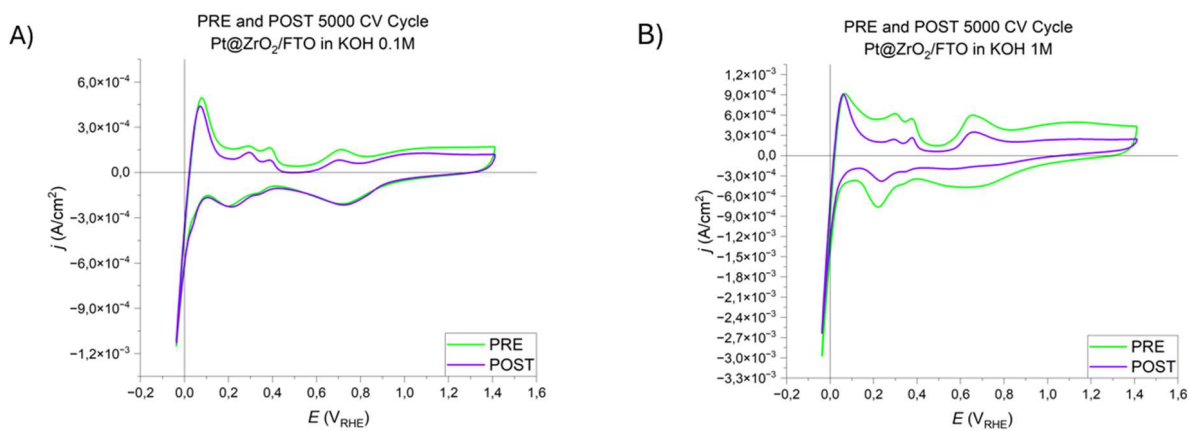

*Figure 5: stability test - comparison between CVs recorded on 50 nm Pt@ZrO2/FTO before the 5000 cycles CVs stress test, PRE, and after the stress test, POST, in (A) 0.1M and (B) 1 M KOH solutions*

*Electrochemical impedance spectroscopy (EIS)*

The new composite electrocatalyst material synthesized by IBSD on FTO was electrochemically characterized by EIS to determine the Charge Transfer Resistance ($R_{ct}$) values at an overpotential of -0.275 V vs. RHE, which is considered a high overpotential for HER.

EIS analysis allows the separation of different contributions: the solution resistance ($R_s$), which is related to the experimental setup and electrolyte conductivity; the capacitive current contribution (C or CPE); and, in some cases, the individual contributions to $R_{ct}$ from the different reaction steps involved in HER under alkaline conditions [50,51]. Specifically, it is sometimes possible to distinguish between the Volmer and Heyrovsky steps (see eq. 1.2 and 1.4 ref [51]), provided that their reaction rates are not too dissimilar. If the difference in reaction rates is too large, fitting the EIS data accurately becomes challenging. For a Pt electrode in an acidic environment, EIS measurements typically exhibit a single semicircle in the Nyquist plot. This semicircle represents an R-C parallel equivalent circuit, where *R* corresponds to the charge transfer resistance and *C* to the double-layer capacitance, modelling the electrochemical system during an electrochemical reaction. In acidic conditions, only a single R-C parallel component is observed because the Volmer step of HER occurs at a significantly higher rate than the Tafel or Heyrovsky steps, making it impossible to resolve their individual contributions. Conversely, in alkaline environments, as reported in the literature [50], the Nyquist plot can reveal the distinct contributions of both the Volmer and Heyrovsky steps, as the Volmer step occurs at a lower rate, allowing their separation.

EIS characterization was performed on both 50 nm Pt@ZrO$_2$/FTO and Pt/FTO. The analysis of IBSD Pt/FTO 50 nm served to validate the previously discussed equivalent circuit models used to fit the experimental EIS data. The results confirmed that, even for a sputtered 50 nm Pt film on an FTO substrate, it is possible to differentiate the contributions of the two reaction steps in an alkaline environment as shown in Figure 6A.

The Nyquist Plot obtained from the electrochemical characterization of IBSD 50 nm Pt/FTO shows the presence of 2 semicircles. In the case of pH 13 (0.1 M KOH) the separation is less defined respect that in the case of pH 14 (1M KOH). In line with a decreasing of Volmer step rate, respect to the Heyrovsky step, as consequence of increase of the pH.

From the fitting of the EIS results, using the equivalent circuit reported in Figure 6C, the $R_{ct}$ value for the two HER reaction steps can be obtained. The first R-C parallel is assigned to Volmer steps (H-UPD; see eq. 1.2 ref [51]), because is the result of high frequency EIS data; the second R-C parallel is related to the Heyrovsky step (HER; see eq. 1.4 ref [51]).

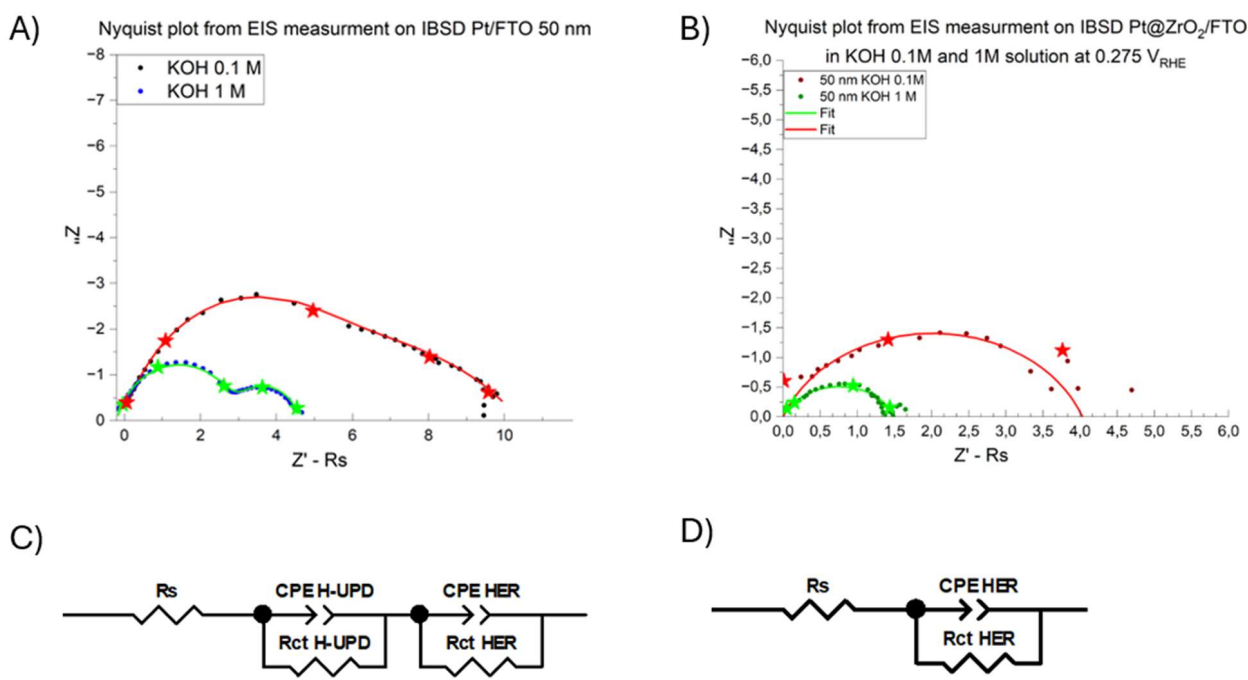

*Figure 6: Nyquist Plot of: (A) IBSD 50 nm Pt/FTO and (B) 50 nm Pt@ZrO₂ in 0.1 and 1 M KOH at the overpotential of -0.275VRHE. The stars represent the data at the frequency 1 Hz, 10 Hz, 100 Hz, 1000 Hz and 10000 Hz. (C) and (D) the relevant equivalent circuits to fit the EIS data.*

In Table 3 are reported the values of the charge transfer resistance obtained from the EIS fitting results using the double parallel RC equivalent circuit (Figure 6C). It is worth note the $R_{ct}$ related to the H-UPD has higher value in both cases, in accordance with the reaction environment that hinder the HER Volmer step.

In the case of Pt@ZrO₂/FTO composite electrocatalyst materials, the synergistic interaction between the noble metal and the oxyphilic ceramic phase is expected to enhance the HER Volmer step rate.

This interaction leads to a scenario similar to that observed in acidic environments for pure noble metals, where the Volmer step becomes too fast to be resolved by EIS.

*Table 3: data obtained from fitting procedure of EIS characterization results of IBSD 50 nm Pt/FTO at -0.275 VRHE*

| KOH (M) | $R_{ct}$ / Ω H-UPD | $CPE\text{-}T$ / S·s$^n$ H-UPD | $CPE\text{-}P$ H-UPD | $R_{ct}$ / Ω HER | $CPE\text{-}T$ / S·s$^n$ HER | $CPE\text{-}P$ HER |
|---|---|---|---|---|---|---|
| 0.1 | 6.87 ± 0.14 | 4.6x10$^{-4}$ ± 6x10$^{-6}$ | 0.78 ± 9x10$^{-3}$ | 3.40 ± 0.19 | 2.3x10$^{-2}$ ± 1x10$^{-3}$ | 0.66 ± 0.03 |
| 1 | 3.15 ± 0.06 | 3.5x10$^{-4}$ ± 3x10$^{-5}$ | 0.82 ± 1x10$^{-2}$ | 1.65 ± 0.07 | 2.1x10$^{-2}$ ± 2x10$^{-3}$ | 0.86 ± 0.03 |

As shown in Figure 6 B, the Nyquist plot from EIS characterization of Pt@ZrO$_2$/FTO in 0.1 and 1 M KOH solutions exhibits only a single semicircle. This behaviour differs from that of IBSD 50 nm Pt/FTO (Figure 6A), where two distinct time constants can be identified. The absence of a second semicircle in the Nyquist plot for Pt@ZrO$_2$/FTO suggests that the Volmer step of HER occurs at a significantly higher rate, making it indistinguishable in the EIS response. Consequently, the equivalent circuit used to fit the EIS data differs from that employed for IBSD 50 nm Pt/FTO, requiring the model reported in Figure 6D.

A key observation from the EIS characterization is that the semicircle corresponding to Pt@ZrO$_2$/FTO is smaller than those obtained for both Pt/FTO 50 nm and Pd@ZrO$_2$/FTO (Figure S2 in Supplementary Information). This indicates a lower charge transfer resistance ($R_{ct}$) for HER on Pt@ZrO$_2$/FTO, which is consistent with the improved electrochemical activity observed in CV measurements. Table 4 and Table 5 summarize the results of the EIS fitting on the above mentioned samples tested in 0.1 M and 1 M KOH. The fittings were performed using ZView 4.0c (Scribner Associates software). It is worth noting the better performance of Pt@ZrO$_2$/FTO in comparison with Pd@ZrO$_2$/FTO, see HER $R_{ct}$ for the tree samples. Finally, Pt@ZrO$_2$/FTO is very similar to Pt/FTO, in term of HER $R_{ct}$, but containing less amount of precious metal.

*Table 4: data obtained from fitting procedure of EIS characterization results at -0.275 V RHE in 0.1 M KOH for different samples*

| Sample | $R_{ct}$ / Ω HER | CPE-T / S·s$^n$ HER | CPE-P HER | $R_{ct}$ / Ω H-UPD | CPE-T / S·s$^n$ H-UPD | CPE-P H-UPD |
|---|---|---|---|---|---|---|
| Pt/FTO 50 nm | 3.40 ± 0.19 | 2.3x10$^{-2}$ ± 1x10$^{-3}$ | 0.66 ± 0.03 | 6.87 ± 0.14 | 4.6x10$^{-4}$ ± 6x10$^{-6}$ | 0.78 ± 9x10$^{-3}$ |
| Pd@ZrO$_2$/FTO 50 nm | 21.87 ± 0.08 | 2.1x10$^{-4}$ ± 6x10$^{-6}$ | 0.78 ± 0.004 | | | |
| Pt@ZrO$_2$/FTO 50 nm | 4.03 ± 0.18 | 1.4x10$^{-2}$ ± 1x10$^{-3}$ | 0.78 ± 0.03 | | | |

*Table 5: EIS data obtained from fitting procedure of EIS characterization results of Pt@ZrO$_2$/FTO at -0.275 V RHE in 1 M KOH*

| Sample | $R_{ct}$ / Ω  HER | CPE-T / S·s$^n$  HER | CPE-P  HER | $R_{ct}$ / Ω  H-UPD | CPE-T / S·s$^n$  H-UPD | CPE-P  H-UPD |
|---|---|---|---|---|---|---|
| Pt/FTO 50 nm | 1.65 ± 0.07 | 2.1x10$^{-2}$ ± 2x10$^{-3}$ | 0.86 ± 0.02 | 3.15 ± 0.06 | 3.5x10$^{-4}$ ± 3x10$^{-5}$ | 0.82 ± 0.01 |
| Pd@ZrO$_2$/FTO 50 nm | 11.91 ± 0.04 | 2.3x10$^{-4}$ ± 8x10$^{-6}$ | 0.83 ± 5x10$^{-3}$ | | | |
| Pt@ZrO$_2$/FTO 50 nm | 1.51 ± 0.04 | 2.1x10$^{-2}$ ± 2x10$^{-3}$ | 0.76 ± 0.02 | | | |

# Conclusions

This study presents a comprehensive electrochemical characterization of Pt@ZrO$_2$ composite electrocatalysts synthesized via Ion Beam Sputtering Deposition method, with a particular focus on their performance in alkaline HER/HOR. The results highlight a significant enhancement in electrochemical activity compared to previous studies on Pd@ZrO$_2$.

The superior performance of Pt@ZrO$_2$/FTO 50 nm over Pd@ZrO$_2$/FTO 50 nm is evident in several key aspects. Firstly, Pt@ZrO$_2$ exhibits a higher current density at -0.05 V vs. RHE, indicating an increased catalytic activity. Secondly, the CV studies reveals enhanced PtO$_x$ reduction and hydrogen underpotential deposition, during the cathodic scan; increase H$_2$ desorption and Pt oxidation, during the anodic scan, thus suggesting a greater number of accessible active sites on the composite electrode. These findings align with the hypothesis that the interaction between Pt and the oxyphilic ZrO$_2$ contributes to an improved catalytic environment, favoring the Volmer step of the HER.

Electrochemical impedance spectroscopy further supports these conclusions. While Pd@ZrO$_2$/FTO displayed a single semicircle in the Nyquist plot [31], indicative of a fast Volmer step that prevents the separation of reaction contributions, Pt@ZrO$_2$/FTO exhibited a smaller semicircle, corresponding to a lower charge transfer resistance. This result confirms that Pt@ZrO$_2$ provides a more efficient pathway for charge transfer, reducing kinetic limitations and enhancing overall HER performance.

In terms of stability, stress tests conducted over 5000 CV cycles in both 0.1 M and 1 M KOH solutions demonstrated that Pt@ZrO$_2$ retained its electrochemical properties in dilute alkaline media. In 1 M KOH, a slight decrease in HER current density was observed, likely due to the partial loss of ZrO$_2$ in more alkaline solution during the 5000 CV cycles, thus leading to an increase in Pt/ZrO$_2$ ration, which, in turn, cause a decrease in the synergist effect of the oxyphilic material.

In conclusion, the integration of Pt within the ZrO$_2$ matrix provides a clear advantage over Pd-based composites. The enhanced charge transfer efficiency, greater stability, and superior HER/HOR kinetics suggest Pt@ZrO$_2$ as a highly promising electrocatalyst for alkaline hydrogen reactions.

## Conflicts of interest

The authors declare no conflicts of interest.

## Acknowledgements

AM acknowledge funding by:
- Funder: project funded under the National Recovery and Resilience Plan (NRRP), Mission 4 Component 2 Investment 1.3 - Call for tender No. 341 of 15.03.2022 of Ministero dell'Università e della Ricerca (MUR); funded by the European Union – NextGenerationEU;
- Award number: project code PE0000021, Concession Decree No. 1561 of 11.10.2022 adopted by Ministero dell'Università e della Ricerca (MUR), CUP D43C22003090001, Project title "Network 4 Energy Sustainable Transition_NEST".

AV acknowledge funding by:
Project funded under the National Recovery and Resilience Plan (NRRP), Mission 4 Component 2 Investment 1.3 - Call for tender No. 1561 of 11.10.2022 of Ministero dell'Università e della Ricerca (MUR); funded by the European Union – NextGenerationEU Project code PE0000021, Concession Decree No. 1561 of 11.10.2022 adopted by Ministero dell'Università e della Ricerca (MUR), CUP D43C22003090001, project title "Network 4 Energy Sustainable Transition_NEST" through the project "Scalable innovative materials and devices for capture and valorisation of CO2 to e-fuels" (eCO2), CUP B53C22004060006

AV and AM are grateful to Piano di Sostegno alla Ricerca 2025 (Linea 2A) – Università degli Studi di Milano.

# Supplementary Information

# Mixed platinum and zirconia powder as electrocatalyst for hydrogen evolution and oxidation reaction

Simone Minelli[1], Manuel Iozzia[2], Claudio Piazzoni[2], Alberto Vertova[*,1,3], Cristina Lenardi[2], Alessandro Minguzzi[1,3,4]

1 Dipartimento di Chimica, Università degli Studi di Milano, Via Golgi 19, 20133 - Milano, Italy.
2 Dipartimento di Fisica, Università degli Studi di Milano, Via Golgi 19, 20133 ` Milano, Italy.
3 INSTM Consorzio Interuniversitario Nazionale per la Scienza e Tecnologia dei Materiali, Via G. Giusti 9, 50121 Firenze, Italy
4 Dipartimento di Energia, Politecnico di Milano, Via Lambruschini 4a, 20156 Milano, Italy

* Corresponding author: alberto.vertova@unimi.it

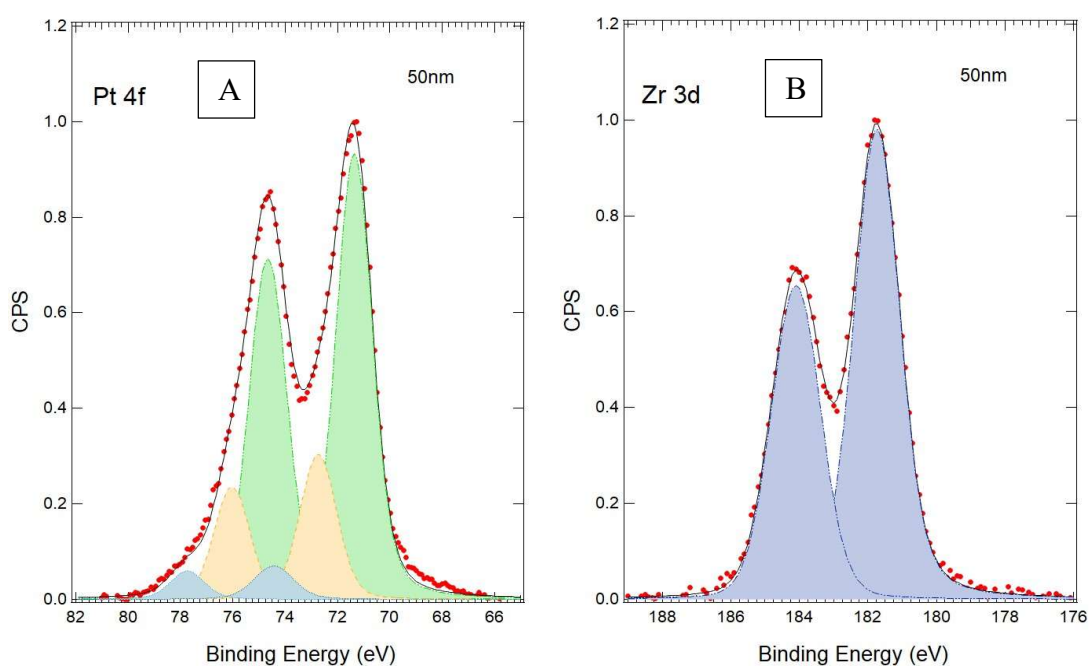

*Figure S1*: XPS detailed scans of Pt 4f (A) and Zr 3d (B) energy regions.

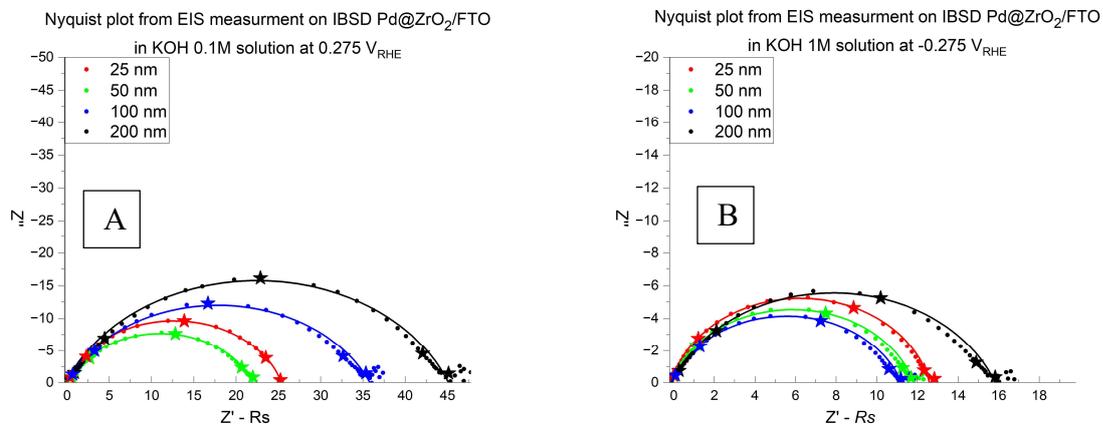

*Figure S2*: Nyquist plot obtained from the EIS characterization made in 3 electrode cell set-up with Pd@ZrO$_2$/FTO as WE in 0.1 M KOH (A) and 1 M KOH (B) solutions at -0.275 V$_{RHE}$. The line represents the fitting results, and the dot are the collected experimental data. The stars represent the data at the frequency 1 Hz, 10 Hz, 100 Hz, 1000 Hz and 10000 Hz